\documentclass[12pt, letterpaper]{article}
\usepackage[top=1in, bottom=1.5in, left=1in, right=1in]{geometry}
\usepackage[utf8]{inputenc}
\usepackage{booktabs}
\usepackage{hyperref}
\usepackage{graphicx}
\usepackage[english]{babel}
\usepackage{longtable}
\usepackage{natbib} 
\usepackage{authblk}
\def\apjl{ApJL}%

\title{Investigating the population of Galactic star formation regions and star clusters within a  Wide-Fast-Deep Coverage of the Galactic Plane}
\author[1]{Loredana Prisinzano\thanks{loredana.prisinzano@inaf.it}}
\author[2]{Laura Magrini\thanks{laura.magrini@inaf.it}} 
\affil[1]{INAF - Astronomical Observatory of Palermo, Italy}  
\affil[2]{INAF - Astrophysical Observatory of  Arcetri, Italy}

\date{November 2018}

\begin{document}

\maketitle

\begin{abstract}
One of the aims of LSST is to perform a systematic survey of star clusters and star forming regions (SFRs) in our Galaxy. 
In particular, the observations obtained with LSST will make a big difference  in Galactic regions that have been poorly studied in the past, 
such as the anticenter and the disk beyond the Galactic center, and they will  have a strong impact in discovering new distant SFRs. 
These results can be achieved by exploiting the exquisite depth that will be
attained if the wide-fast-deep (WFD) observing strategy of the main survey is also adopted for the Galactic plane,
in the {\it g}, {\it r}, and {\it i} filters. 
\end{abstract}
\section{White Paper Information}
\begin{footnotesize}Francesco Damiani (INAF-OAPa-Italy),  Germano Sacco (INAF-OAA-Italy), 
Rosaria Bonito (INAF-OAPa-Italy), Laura Venuti (Cornell University, Ithaca, NY-US),  Giada Casali (INAF-OAA-Italy), Veronica Roccatagliata (INAF-OAA-Italy), Sofia Randich (INAF-OAA-Italy), Laura Inno (INAF-OAA-Italy), Tristan Cantat-Gaudin (Universitat de Barcelona, Spain), Dante Minniti (Universidad Andres Bello, Santiago, Chile), Angela Bragaglia (INAF-OABo-Italy), Degli Innocenti Scilla (Pisa University-Italy), Pier Giorgio Prada Moroni (Pisa University-Italy), Emanueli Tognelli (Pisa University-Italy), Antonio Sollima (INAF-OABo-Italy),  Antonella Vallenari (INAF-OAPd-Italy), Mario G. Guarcello (INAF-OAPa-Italy), Sergio Messina (INAF-OACt-Italy), Giuseppina Micela (INAF-OAPa-Italy), Salvatore Sciortino (INAF-OAPa-Italy), Alessandro Spagna (INAF-OATo-Italy), Nicoletta Sanna (INAF-OAA-Italy),  Joel Kastner (Rochester Institute of Technology, US), Alexandre Roman (Universidad de La Serena - Chile), Robert Szabo (MTA CSFK, Konkoly Observatory, Hungary), Bruno Dias (ESO Santiago, Chile), Eric Feigelson (Pennsylvania State University),
Rob Jeffries (Keele University, UK), Eileen Friel (Indiana University), John Stauffer (IPAC, California Institute of Technology), Mathieu Van der Swaelmen (INAF-OAA-Italy), Rosanna Sordo (INAF-OAPd-Italy), Diego Bossini (INAF-OAPd-Italy)
\end{footnotesize}

\begin{enumerate} 
\item {\bf Science Category:} Mapping the Milky Way
\item {\bf Survey Type Category:} Wide-fast-deep
\item {\bf Observing Strategy Category:} 
a specific pointing or set of pointings that is (relatively) agnostic of the detailed observing 
	strategy or cadence, (e.g., a science case enabled by very deep precise multi-color 
	photometry)
\end{enumerate}

\clearpage

\section{Scientific Motivation}

\begin{footnotesize}

Interest in star clusters is driven by several scientific topics, e.g.: star clusters are laboratories for understanding the physical mechanisms active in stars and are tracers of Galactic chemical evolution  thanks to their wide range in ages and distances, both in the disk (open clusters) and in the halo (globular clusters). Moreover, the SFRs are the place where we can see star formation in action.
Our knowledge of  young clusters and  SFRs is currently limited to nearby regions, while LSST observations will allow us to discover SFRs at distances 7 times farther than those achievable  even with the deepest wide-area  surveys, e.g. PanSTARRS1. This will allow us, for example, 
 to discover the low-mass star population (down to or even below 0.5 M$_\odot$) of even quite extincted  SFRs (A$_V\sim5$) 
in the volume shell from $\sim$1 kpc up to $\sim$10 kpc, which would remain unexplored 
by other surveys.
Regarding the oldest clusters, LSST will allow us to detect them in the outer Galactic disk, not only located at high latitudes, like the old clusters already known, but also in the Galactic Plane. 
\subsection{Star forming regions}
We propose a new approach to select  low mass young stars and to  discover SFRs based on the 
$g-r$ vs. $r-i$  colour-colour diagram (CCD) that will be obtained with the LSST co-added depth.
This method is alternative and complementary to that based  
on the variability properties of young stars, 
mentioned in the {\it Science-Driven Optimization of the LSST Observing Strategy}
\citep{lsst17}.
 The approach is based on the co-exploitation of several properties of M-type stars during 
 the pre-main sequence (PMS) phase \citep[][Venuti et al. 2018, in press]{dami18,pris18}, as will be described in the following section.
The numeric fraction of M-type stars in the
Galactic disk
is $\sim$84\%   \citep{lada06} and they represent the largest component of the stellar populations. They are  crucial to
 investigate fundamental aspects of star formation (SF),
 such as the Initial Mass Function (IMF), the modes and timescales of SF and
PMS evolution, or the mechanisms of cluster dispersal \citep{bric07}.
A quantitatively measure the size/mass of young clusters
as a function of galactocentric distance and/or Fe/H might help provide new insights into
how cluster formation happens and what fraction of newly formed
stars are in big clusters vs. more isolated birth sites.
The most common solar-metallicity M-dwarfs (dM) 
can be unambiguously      selected in  CCD
involving the {\it r} magnitude, such as for example the {\it g-r} vs. {\it r-i} diagram.
 In fact, due to the strong TiO opacity, the {\it r}-band flux of dM stars is 
depressed \citep{west11} and the CCDs shows the well known "elbow"  where the dM locus 
deviates from the earlier-type (hotter/earlier than M) stars \citep{fuku96},   nearly parallel
to the reddening vector. 
PMS M-type stars share the same spectral properties as dM stars but are characterized 
by a significantly different gravity, which increases as function of age during the  evolution from the PMS to the Zero Age MS.
In fact, during the PMS phase, Hayashi evolutionary tracks are almost vertical with an excursion 
in luminosity of  $\sim$1.5 dex. {\bf  This implies    that  
PMS M stars can be detected at a distance up to a factor 5  larger than   dM stars}.
In this context, the selection of young M-type stars in distant SFRs
is favoured by a significantly higher stellar density with respect to the more dispersed
 and uniformly distributed dM stars in the field, that,
  can only be detected up to  a much lower distance. 
Due to their larger intrinsic luminosity, PMS M-type stars
will sample a far larger volume  than MS stars at similar apparent magnitudes.
Our ability to detect PMS M-type stars strongly depends on extinction along the
line of sight. The Galactic plane has a patchy structure including many crowded regions,
where the final LSST co-added depth is strongly affected  by confusion limits,
but  also many obscured  regions, where the optical  source detection
has been hampered by the extinction. 
The LSST final co-added  depth that would be achieved  by adopting the WFD observing strategy also in the most
 unexplored
regions of the Galactic plane, will enable us to discover the very low mass population of the 
SFRs at distances that are inaccessible even with deep IR photometric surveys.
In fact, the latter only allow us to discover samples biased in favour of young stars with circumstellar disks, 
while the proposed approach enables us to select larger samples
of  PMS M-type stars, irrespective of the presence of a disk. 
The discovery of new SFRs with LSST data by adopting this method  is supported by the observational
findings  showing that 
SFRs mostly host  1-2\,Myr
old members \citep{hill97,pall99,venu18},  which is the age range for which the proposed method is the most effective.
In fact at 1-2\,Myr, theoretical isochrones of M-type stars predict luminosities larger than those
predicted for older ages. 

\subsection{Star clusters}

Open clusters represent an important fraction of the stellar population of our Galaxy and they are mainly located in the Galactic thin disc. 
Thus, they can be used to trace the spatial metallicity distribution inside the Galactic disk \citep[see, e.g.][]{janes79, twarog97, friel02, magrini17, magrini18}. Complementary to the study of the overall metallicity distribution, the abundance ratios of several elements can provide insightful information both on the SF history in the disk and on the nucleosynthesis processes, production sites and timescales of enrichment of each element. For a complete review of the role of open clusters see \citet{friel13}. 

In this white paper, we focus our analysis on the possibility to {\bf detect star clusters in the outer disc at any height above the Galactic Plane, including those at low latitudes}. On one hand, in the anticenter  most of the known open clusters at distances $>$12~kpc are located at high altitudes above the Galactic plane. Many hypotheses have been proposed to explain this apparent lack of star clusters in the Galactic outskirts at low Galactic heights: i) this might be due to observational biases caused by the higher reddening towards the plane; ii) the Galaxy might be flared and thus the clusters might trace the flare; iii) their location might be due to the effects of stellar migration, particularly effective at the corotation radius. On the other hand, almost nothing is known about clusters located on the other side of the Galaxy, and our knowledge is limited to clusters located in a small region.
The aim of LSST is {\bf the detection of star clusters together with their full characterization}: from the upper main sequence to its faint end, from the evolved red giant stars to the white dwarf population. To this purpose, a range of  magnitudes, together with accurate proper motions, are needed.
The availability of a consistent set of data from the dwarfs to the giants in a cluster allows to calibrate stellar models on samples of coeval objects, in particular for free parameters still not fully constrained \citep[e.g., helium-to-metal enrichment ratio, the efficiency of super-adiabatic convection, the convective core overshooting parameter][]{gennaro10, tognelli12, prada12}. 
Examples of the kind of work we can do with LSST are shown in \citet{richer08} and in \citet{bab18}, where cluster HR diagrams are cleaned with proper motion. 
To understand the limits of the detection of stellar clusters with the different cadences, we considered a cluster with a typical age (0.3 Gyr) and metallicity (solar), and we have located it at several distances and with several extinction (see Fig.\ref{iso}, Bottom, right panel). 
 The number of open clusters observed by LSST will be huge. Consider, for instance,  the recent works by \citet{castro18} and \citet{cantat18}, which,  with  Gaia DR2 data,  discover 30 new OCs in the Solar neighborhood and 28 in the direction of Perseus, respectively, i.e., in regions in which the census of clusters was considered complete, increasing by 20\% their number (see Fig.\ref{iso}, Bottom, left panel).
 The number of known open clusters in the  Dias catalogue is about $\sim$2200 \citep[][in the new version of 2016]{dias02}. 
The completeness  is a decreasing function of the distance, reaching few percent in the outer regions, LSST will allow us to discover from $\sim$400 to more than 2000 new clusters in the Galactic plane. 
In addition, the {\bf population of globular clusters} (GCs), which  represent the oldest populations of our Galaxy, remains unexplored in regions such as the Galactic plane and bulge.
The catalog of Galactic GCs originally contained 156 members \citep{harris96}, but new GCs have been recently discovered  in the Galactic plane and bulge \citep[e.g][]{minniti11, moni12, minniti17a, minniti17b, minniti18}. 
The difficulties related to the discovery of clusters in the Galactic plane, as the stellar density combined with (differential) interstellar reddening, are reduced with the deep photometry of LSST. These new GCs, detected as high density regions in LSST maps, would be confirmed as globular cluster candidates by their CMDs. The new globular cluster candidates would exhibit a variety of extinctions, and distances, that can also be measured using the LSST multicolor-photometry, and variable stars like RRLyrae. 
 The discovery and characterization of these new GCs is very important for studies on the formation and evolution of the MW, on the age and chemical composition of the oldest stars, on the dynamics of stellar systems, on the interstellar medium, and on Galactic structure and stellar evolution.

\end{footnotesize}

\newpage

\begin{figure}[h]
 \centering
 \includegraphics[width=11cm]{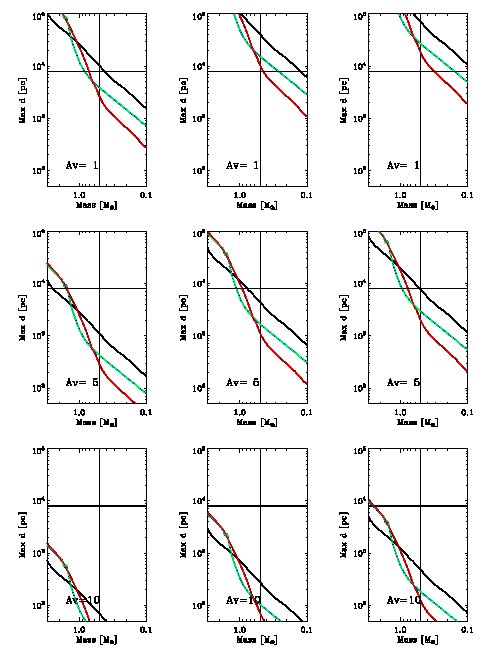}
\includegraphics[width=8cm]{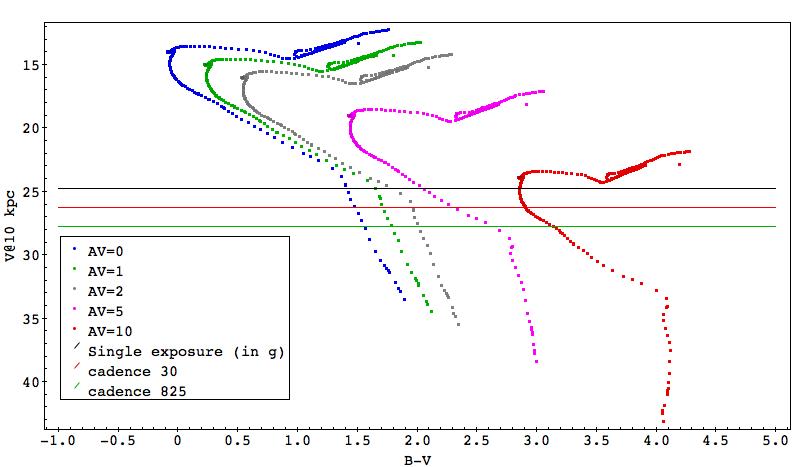}
\includegraphics[width=6cm]{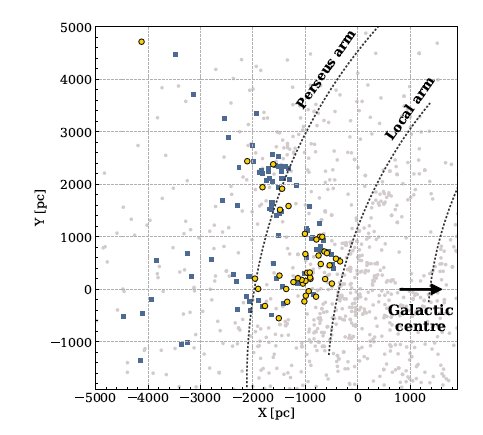}

\caption{\tiny{Top: Maximum distances as a function of stellar mass in the PMS phase (1-10\,Myr, black and green lines, respectively) 
and at 100\,Myr (red lines),
at different extinctions,  assuming A$_V$=1, 5 and 10 mag. Horizontal solid line indicates the distance 
of the Galactic Center, while the vertical one  is drawn at 0.5\,M$_\odot$. The values are given assuming the
PanSTARRS1 (left panels), the "30-visits-maglim" (central panels) and the "WFD-survey-maglim" (right panels)
 5\,$\sigma$ stacked depths.
Bottom:  {\em Left} : Theoretical CMD of a cluster of 0.3 Gyr in the outer disc/beyond the Galactic center at a distance of 10 kpc with five reddening values. 
{\em Right}: Locations of open clusters on the projected Galactic plane, for both previously known  (grey dots) and new COIN-Gaia open clusters (yellow dots) from \citet{cantat18}.}}
\label{iso}
 \end{figure}


\vspace{.6in}

\section{Technical Description}

\subsection{High-level description}
\begin{footnotesize}
Add the currently excluded Galactic Plane  region to the uniform (wide-fast-deep) cadence.
\end{footnotesize}

\vspace{.3in}

\subsection{Footprint -- pointings, regions and/or constraints}
\begin{footnotesize}
Including the regions of the Plane, above and below the Galactic Plane (10$^o$). 
\end{footnotesize}

\subsection{Image quality}
\begin{footnotesize}
No constraints.
\end{footnotesize}

\subsection{Individual image depth and/or sky brightness}
\begin{footnotesize}
No extra constraints beyond the standard lunar avoidance for WFD.
\end{footnotesize}

\subsection{Co-added image depth and/or total number of visits}
\begin{footnotesize}
Uniform WFD cadence.
\end{footnotesize}

\subsection{Number of visits within a night}
\begin{footnotesize}
No constraints.
\end{footnotesize}

\subsection{Distribution of visits over time}
\begin{footnotesize}
Uniform cadence over 10 years to allow good proper motion and parallaxes measurement.
\end{footnotesize}

\subsection{Filter choice}
\begin{footnotesize}
Standard WFD set of filters, with priority to {\it g}, {\it r}, and {\it i} filters. 
\end{footnotesize}

\subsection{Exposure constraints}
\begin{footnotesize}
Standard WFD exposure.
\end{footnotesize}

\subsection{Other constraints}
\begin{footnotesize}
\end{footnotesize}

\subsection{Estimated time requirement}
\begin{footnotesize}
The estimate of the time requirement of a WDF survey of the Plane will amount to 2\% of the duration of the 
whole survey. 
\end{footnotesize}

\vspace{.3in}

\begin{table}[ht]
    \centering
    \begin{tabular}{l|l|l|l}
        \toprule
        Properties & Importance \hspace{.3in} \\
        \midrule
        Image quality &   1  \\
        Sky brightness &  1 \\
        Individual image depth & 3  \\
        Co-added image depth &  1 \\
        Number of exposures in a visit   & 3  \\
        Number of visits (in a night)  &  3 \\ 
        Total number of visits &  1 \\
        Time between visits (in a night) & 3 \\
        Time between visits (between nights)  & 1  \\
        Long-term gaps between visits & 2\\
        Other (please add other constraints as needed) & \\
        \bottomrule
    \end{tabular}
    \caption{{\bf Constraint Rankings:} Summary of the relative importance of various survey strategy constraints. Please rank the importance of each of these considerations, from 1=very important, 2=somewhat important, 3=not important. If a given constraint depends on other parameters in the table, but these other parameters are not important in themselves, please only mark the final constraint as important. For example, individual image depth depends on image quality, sky brightness, and number of exposures in a visit; if your science depends on the individual image depth but not directly on the other parameters, individual image depth would be `1' and the other parameters could be marked as `3', giving us the most flexibility when determining the composition of a visit, for example.}
        \label{tab:obs_constraints}
\end{table}

\subsection{Technical trades}
\begin{footnotesize}
In our proposal the primary science goals are reached with the increased depth of the co-added images and also  with the repeated observations with a long time baseline in order to measure proper motions and parallaxes. 
Considering the distribution of visits: for our science cases a uniform coverage over the whole survey 
is desirable to have a long baseline for measuring parallaxes and proper motions.  

\end{footnotesize}

\section{Performance Evaluation}
\begin{footnotesize}
{\bf Star forming regions:} For PMS M-type stars, the key interest regarding the observing strategy 
of the survey is mainly the photometric co-added depth. 
Since the method is based on the {\it g}, {\it r}, and {\it i} CCD, 
the metrics is dependent on the limiting magnitudes in these three bands. In fact, 
the maximum distance to which PMS stars can be detected is basically set
by the filter with the lowest sensitivity. This distance depends on stellar mass, age and extinction.
Fig.~\ref{iso} shows the maximum distance as a function of the stellar masses including the
range of M-type stars (M$\leq$0.5 M$_\odot$). The results were obtained by assuming the 5\,$\sigma$ 
stacked point-source depths of the simulated ten-year survey, roughly estimated from Figure 3.2 of
LSST Science Collaborations and LSST Project (2009, LSST Science Book,
Version 2.0, arXiv:0912.0201), that are
{\it g}=26.3, {\it r}=26.5, {\it i}=26.3, for the 30 visits for each filter planned in the Galactic plane and
{\it g}=27.5, {\it r}=27.7, {\it i}=27.0, for the 80, 180 and 180 visits, respectively, in the {\it g}, {\it r}, and {\it i} filters, as planned in the main survey in the
regions outside the Galactic plane. For comparison, we also estimated the maximum distances by considering 
the 3$\pi$ stack 5$\sigma$  depths of PanSTARRS1 magnitudes {\it g}=23.3, {\it r}=23.2, {\it i}=23.1.
 We will refer to the first set of limiting magnitudes as the "30-visits-maglim" and to the second  one
 as the "WFD-survey-maglim"
 
The distances have been obtained by  considering the absolute magnitudes in the same filters
in the  1, 10 and 100\,Myr  isochrones computed by \citet{togn18}, assuming  A$_V$=1, 5 and 10\,mag,
i.e. covering the cases of low , high,  and very high reddened regions \citep{gree18}. 
Then a figure of merit can be deduced from the volume  that can be mapped with PMS M-type and dM stars
at a given extinction, as a function of stellar masses.  The values of the maximum distances obtained
assuming  the "30-visits-maglim" and the "WFD-survey-maglim" depths are 
  shown in Table\,\ref{figureofmeritall}.
  For comparison,
  the analogous values that will be achievable with PanSTARRS1 are also given.

 The exquisite LSST WFD co-added depth will have a strong 
impact in the context of the SFR studies.
In fact, it will make possible   to detect the very low mass
population (down to 0.1M$_\odot$) of highly embedded (A$_V\sim$5)  SFRs, up to $\sim$1\,kpc,
 well beyond what could be achieved with the "30-visits-maglim" (up to 0.67 kpc);
or the PanSTARRS1 limit (0.17 kpc). Instead, in case of low reddening (A$_V\sim$1),
it will be possible to discover the entire cluster population
of very young stars down to 0.1\,M$_\odot$  up to 10\,kpc,
involving a very large unexplored  volume (e.g. in the direction of Vela-Puppis)
that would be inaccessible with  "30-visits-maglim"
depth. 
The difference in depth between the LSST WFD and the 30-visits  strategy
 (about 1.2 magnitude) corresponds to about a factor 1.7 in distance. 
 Since the number of star clusters roughly increases with d$^2$,
 we define the figure of merit (FoM) as the relative fraction of detected clusters (FoM=N/N$_{\rm WDF}$).
Therefore, {\bf for the  WFD co-added depth  FoM=1, while for the 30-visits co-added depth  FoM=0.34}. 
 For comparison, the difference in depth between the LSST WFD strategy and PanSTARRS1
 (about 4.2 magnitude) corresponds to about a factor 7 in distance   and therefore for the PanSTARRS1
 observations  FoM=0.02.  {\bf This means that by adopting the WFD observing strategy in the Galactic plane,
  we will be able to discover a number of clusters at least a 
  factor 3 larger than  the one that would be discovered by adopting the 30-visits observing strategy.  }
 \begin{table*}
\caption{Maximum distances at 0.5 and 0.1\,M$_\odot$ at 1, 10 and 100\,Myr and for A$_V$=1, 5 and 10\,mag assuming
PanSTARRS, 30-visits and WFD-survey   depths.
The max distances at 1 Myr for a given Av and M* are highlighted in boldface.
 \label{figureofmeritall}}
\centering
\begin{tabular} { c c c c c c }  
\hline\hline
A$_V$ &  Age        & Mass         & max distance  & max distance & max distance \\
 mag    &   [Myr]    & [M$_\odot$]   &  [kpc] & [kpc] & [kpc] \\
     &     &    & PanSTARRS1 &  30-visits & WFD-survey \\
\hline
  1 &    1&0.5 &{\bf 10.33} &{\bf 41.12} &{\bf 71.46}\\
  1 &   10&0.5 &  3.88 & 15.43 & 26.81\\
  1 &  100&0.5 &  2.61 & 10.40 & 18.07\\
  5 &    1&0.5 &{\bf  1.12} &{\bf  4.46} &{\bf  7.75}\\
  5 &   10&0.5 &  0.42 &  1.67 &  2.91\\
  5 &  100&0.5 &  0.28 &  1.13 &  1.96\\
 10 &    1&0.5 &{\bf  0.07} &{\bf  0.28} &{\bf  0.48}\\
 10 &   10&0.5 &  0.03 &  0.10 &  0.18\\
 10 &  100&0.5 &  0.02 &  0.07 &  0.12\\
  1 &    1&0.1 &{\bf  1.54} &{\bf  6.14} &{\bf 10.68}\\
  1 &   10&0.1 &  0.72 &  2.88 &  5.01\\
  1 &  100&0.1 &  0.27 &  1.09 &  1.89\\
  5 &    1&0.1 &{\bf  0.17} &{\bf  0.67} &{\bf  1.16}\\
  5 &   10&0.1 &  0.08 &  0.31 &  0.54\\
  5 &  100&0.1 &  0.03 &  0.12 &  0.21\\
 10 &    1&0.1 &{\bf  0.01} &{\bf  0.04} &{\bf  0.07}\\
 10 &   10&0.1 &  0.00 &  0.02 &  0.03\\
 10 &  100&0.1 &  0.00 &  0.01 &  0.01\\
\hline
\end{tabular}
\end{table*}

{\bf Star clusters: } For the star clusters located in the regions with the high reddening, we would benefit to be included in the main WFD survey because with the 30 visits (planned for the Galactic plane) we will arrive to observe only the turnoff and red giant clump, with a consequently lower accuracy in deriving the cluster parameters (age, metallicity, reddening via isochrone fitting).   
We are also performing a more detailed analysis of the extinction through the disk using the new dust map of \citet{gree18} to derive the limits in magnitudes for clusters in the disk at different galactic coordinates and distances.    
A FoM to evaluate the advantages of the WDF survey with respect to the 30 visits of the Galactic disk can be estimated on the basis of  the limiting magnitude that can be reached and which part of a cluster CMD can be observed given this limiting magnitude.  We have considered that the WDF survey should arrive $\sim$1.2 mag deeper than the 30 visits, and for instance, it would allow a better characterization and isochrone fitting of the most extinct ones. 
Similarly to the star formation regions, since the number of clusters scales with d$^2$,  the WFD observing strategy in the plane will allow us to enlarge the number of old star clusters that we can discover and characterize of at least a factor 3  with respect to the 30-visits observing strategy.
If we consider a FoM=1 for the WDF Survey of the Galactic Plane, the lower magnitude limit due to only 30 visits, would imply, as for the star forming regions, a FoM=0.34. 

In addition, another important aspect to be considered is the measurements of proper motions, fundamental to identify cluster members. 
Proper motions can be measured by LSST with an accuracy of 0.1 mas yr$^{-1}$ in the magnitude range 16$<$r$<$20, and the accuracy linearly decreases until 1 mas yr$^{-1}$ for 20$<$r$<$24.
Typical values of the proper motion for clusters within few kpc are 1-5 mas yr$^{-1}$. These numbers translate into 0.5-2 mas yr$^{-1}$ for clusters at 10 kpc. 
For r=22 (corresponding to the upper main sequence for an extinct cluster or to the intermediate MS for a less extinct one) we will achieve the  sufficient precision required to distinguish a cluster member from field stars. A more frequent cadence will help to separate proper motions from parallaxes and to improve the final accuracy. 
A FoM to evaluate the advantages of the deep-wide-fast survey with respect to the 30-visits survey of the Galactic plane is given by the final accuracy on the proper motions and ability to separate them from parallaxes, an operation that requires a long temporal baseline.  
Following \citet{ive08}, the expected proper motion and parallax errors for a 10-year long baseline survey, as a function of apparent magnitude, are 0.6 mas and 0.2 mas yr$^{-1}$ for r=21, respectively, 0.8 mas and 0.3 mas yr$^{-1}$ for r=22, 1.3 mas and 0.5 mas yr$^{-1}$ for r=23 and finally 2.9 mas and 1.0 mas yr$^{-1}$ for r=24.   
However, the accuracy on these quantities are lower with a shorter temporal baseline, and with a lower number of visits. For instance, the proper motion errors increase of a factor 5 and the parallax errors of a factor 2 with a shorter baseline of three years.  Considering a FoM=1 for the WDF survey of the Plane, and increasing errors of a factor 5 on proper motions due to shorter temporal baseline and/or lower number of visits would imply a FoM=0.2. 

{\bf Considering both the requirements on the magnitude limit to fully characterize a star clusters (FoM=0.34) and on the proper motion to identify its member stars (FoM=0.2), we will have an overall FoM=0.068 accepting a strategy of only 30 visits with respect to the WDF strategy (FoM=1). }

\end{footnotesize}

\vspace{.6in}

\section{Special Data Processing}
\begin{footnotesize}
Standard LSST Data Management pipelines are requested
\end{footnotesize}

\section{Acknowledgements}
 \begin{footnotesize}
 \end{footnotesize}

\section{References}
\bibliography{bibdesk}

\end{document}